\documentclass[aps,twocolumn,prb,showpacs,amsmath,amssymb,floatfix]{revtex4}
\usepackage{graphics}
\usepackage{epsf,graphicx,epsfig}
\usepackage{color}
\newcommand{\bk}{\mathbf{k}}
\newcommand{\dg}{^\dagger}

\begin{document}

\title{Quasiparticle interference in an iron-based 
superconductor}
\author{S.~Sykora$^{1,2}$, and Piers~Coleman$^{1}$}

\affiliation{
  $^{1}$Center for Materials Theory, Rutgers University,
  Piscataway, New Jersey 08854, USA \\
$^{2}$IFW Dresden, Institute for Theoretical Solid State Physics, P.O. Box 270116, 
D-01171 Dresden, Germany
} 

{\today}



\begin{abstract}
We develop a model for the effect of a magnetic field on quasiparticle interference in 
an iron-based superconductor. Recently, scanning tunneling experiments have been 
performed on Fe(Se,Te) to determine the relative sign of the superconducting gap from the 
magnetic-field dependence of quasiparticle scattering amplitudes. Using a simple two-band
BCS model, we study three different cases of scattering in a spin-split spectrum. The 
dominant effect of a magnetic field in iron-based superconductors is caused by the 
Pauli limiting of conduction electrons. Thereby time reversal odd scattering is induced 
which enhances the sign-preserving and depresses the sign-reversing peaks in 
the quasiparticle interference patterns.
\end{abstract}

\maketitle

\section{Introduction}

The discovery of iron-based layered pnictide and chalcogenide
superconductors\cite{KW2008} with superconducting (SC) transition
temperature as high as 55K (Ref.~\cite{RE2008}) has generated enormous
interest in the physics of these materials. Similar to the cuprates, the
iron-based superconductors are highly two-dimensional and superconductivity occurs in
close proximity to anti-ferromagnetic order,\cite{LD2009} leading
several groups to propose that the pairing is driven by
anti-ferromagnetic spin fluctuations.\cite{MS2008}-\cite{CT2009}

The
first step in identifying the pairing mechanism is to investigate the
structure of the SC-gap function, which describes the strength and
quantum mechanical phase of electron pairs in momentum (${\bf k}$)
space. While the SC gap function of conventional phonon-mediated
superconductors has the same phase throughout momentum space (s-wave
symmetry), that of spin-fluctuation mediated superconductors is
expected to exhibit a sign reversal between those Fermi momenta
connected by characteristic wave vector ${\bf Q}$ of the spin
fluctuations.\cite{KA2001} In $d$ wave superconductors, the nodal
planes of the order parameter intersect with the Fermi surface,
leading to gapless quasiparticle (QP) excitations that can be detected
thermodynamically and by low energy probes. However, if the sign
reversal occurs between disconnected hole and electron pockets
(compare Fig.~\ref{Fig_FS}) resulting in an ``$s_{\pm}$ symmetry,''\cite{MS2008,KU2008} the relative sign of SC gap must be determined by
phase-sensitive experiments such as Josephson junctions
experiments\cite{TK2000} or composite SC loops.\cite{CC2009} An
alternative technique is scanning tunneling microscopy (STM) which
determines the dispersion of 
quasiparticle states from the quasiparticle interference
(QPI) patterns induced by impurity
scattering\cite{HO2002}-\cite{HA2009}. Such experiments,
performed in an external magnetic field, offer the capability 
of probing the phase of the superconducting order parameter by
detecting a field enhancement of the sign-preserving scattering that results
from the sensitivity of QPI to the coherence factors associated with
impurity scattering.

The recent detection of a field enhancement of quasiparticle
scattering between the hole pockets of the iron chalcogenide superconductor
Fe(Se,Te) has been interpreted\cite{HA2010} as evidence for the $s_{\pm}$ pairing
scenario.  
However, there are certain differences between the field effect
in iron-based and cuprate superconductors that raise questions about
this interpretation.  For example, Fe(Se,Te) is a Pauli-limited
superconductor where the coupling of the field to the conduction
electrons becomes important. In this material, 
the Zeeman splitting is a large fraction of
the superconducting gap size,
$\mu_B B / \Delta \approx 0.4$, generating new components to the
field-induced scattering that have been hitherto neglected in models
of quasiparticle interference. 

In the cuprates, quasiparticle scattering 
in external magnetic fields is strongly affected by vortices. However,
STM measurements on Fe(Se,Te) did not observe appreciable correlation
in space between the location of vortices and the magnitude of field-induced
change in the QPI intensity.\cite{HA2010} These results suggest that in
Pauli-limited superconductors the role of vortices is much less important 
compared  to homogeneously distributed field-induced scatterers created 
by Zeeman splitting. One of the key arguments advanced in
the STM measurements of Fe(Se,Te), is that an applied magnetic field
will, in general, enhance the time-reverse odd components of the
scattering that lead to sign-preserving scattering.   How effective
are these arguments in the presence of significant Pauli limiting
effects, and to what extent can we attribute the observed field enhancement
of the sign-preserving scattering to 
the Zeeman splitting?

Motivated by the above considerations, this paper develops a
phenomenological model for the field-dependent QPI in iron-based
superconductors, taking into account the Zeeman splitting of the
quasiparticle dispersion. Using a simple two-band BCS model for an
unconventional $s$-wave superconductor, we calculate the STM conductance
ratio for different types of disorder in an external magnetic
field at which we neglect the scattering off vortices. In our model, 
we assume an $s_{\pm}$-wave symmetry for the
superconducting order parameter which means that the size of the
superconducting gap is isotropic along the hole and electron pockets.
Our study confirms that Zeeman splitting provides a strong source of
time-reversal-odd, sign-preserving scattering. We also find that if
resonant or magnetic scatterers are present in the superconducting
material, the spin split spectrum leads to an enhancement of sign
preserving scattering. QPI patterns for non-magnetic and magnetic impurities without
Zeeman field have been computed theoretically by a number of works using more realistic band 
structures.\cite{WZL2009}-\cite{AK2010} 

The paper is organized as follows. In Sec.~II~A, we briefly resume the
basic concepts behind some types of scatterers with different time
reversal symmetry properties and their coherence factors. The BCS
model for an iron-based  superconductor will be described in Sec.~II~B.
After a short discussion about calculating the tunneling conductance
from local density of states in Sec.~III, we present our results of the
numerical evaluation in Sec.~IV. We conclude in Sec.~V.

\section{Phenomenological model}

\subsection{Coherence factors}

An  STM measurement probes  quasiparticle interference (QPI) by 
observing Friedel
oscillations in the tunneling density of states induced by
impurity scattering.   The Fourier transform of these fluctuations in
the density of states provides information about the
rate at which quasiparticles are scattered by impurities. 
In general, the rate of impurity scattering
is proportional to the coherence factors associated with the
underlying scattering mechanism.  
By measuring the momentum dependence of the scattering, it becomes
possible to extract phase sensitive information about the underlying
order parameter.

Quasiparticles in a superconductor are a
coherent superposition of electrons and holes. Coherence factors 
describe the difference between the scattering rates of a SC quasiparticle and a bare 
electron off a given scatterer.\cite{TI2004} 
In general, 
the scattering rate of an electron from initial momentum ${\bf k}$ to final
momentum ${\bf k}'$ is determined by an energy-dependent $t$-matrix 
$\hat{T}({\bf k},{\bf k}',\omega)$. The $t$-matrix, which depends 
on the quasiparticle spin degrees of freedom 
is most conveniently described using four-component
Nambu notation\cite{BW1963},\cite{BVZ2006}. For a single impurity 
with scattering potential $\hat{U}({\bf k},{\bf k}')$ in momentum space, the corresponding 
$t$-matrix exactly accounts for multiple scattering off that 
impurity.\cite{BVZ2006},\cite{HVW1986} In first-order perturbation theory (Born approximation)
 the $t$-matrix is equal to $\hat{U}({\bf k},{\bf k}')$. 
In four-component Nambu notation the electron field 
is described by a 
four-component spinor containing particle and 
hole operators,\cite{BW1963}
\begin{equation}
\label{1}
\Psi_{\mathbf{k}}^\dag =
\left(
c_{\mathbf{k},\uparrow}^\dag \;
c_{-\mathbf{k},\downarrow}^\dag \;
c_{-\mathbf{k},\downarrow} \;
-c_{\mathbf{k},\uparrow}
\right) ,
\end{equation}
where $\mathbf{k}$ is the momentum.

We now review the connection between 
coherence factors for scattering in a BCS superconductor and the 
time-reversal symmetry of the scattering process.
Consider a scattering
process described by the $t$-matrix,
\begin{equation}
\label{scparticles}
\hat T = \sum_{\mathbf{{k}},\mathbf{k}'}\Psi_{\mathbf{k}}^\dag \, \hat{T}_{\bk ,\bk '} \, \Psi_{\mathbf{k}'},
\end{equation}
where the Nambu-spinors $\Psi_{\mathbf{k}}^\dag$ and $\Psi_{\mathbf{k}'}$ account for different 
particle states according to Eq.~\eqref{1}. 
In the superconducting state, the $t$-matrix determines the Friedel
oscillations
in the local density of states that are probed in
an STM experiment. 
To determine the matrix elements for 
quasiparticle scattering, we apply a Bogoliubov transformation to the particle 
spinors in Eq.~\eqref{scparticles}, leading to 
\begin{equation}
\label{Bogoliubov}
\left(
\begin{array}{c}
c_{\mathbf{k},\uparrow} \\
c_{-\mathbf{k},\downarrow} \\
c_{-\mathbf{k},\downarrow}^\dag \\
-c_{\mathbf{k},\uparrow}^\dag
\end{array}
\right) = 
\left(
\begin{array}{cccc}
u_{\mathbf{k}} & 0 & v_{\mathbf{k}} & 0 \\
0 & u_{\mathbf{k}} & 0 & v_{\mathbf{k}} \\
-v_{\mathbf{k}} & 0 & u_{\mathbf{k}} & 0 \\
0 & -v_{\mathbf{k}} & 0 & u_{\mathbf{k}}
\end{array}
\right)
\left(
\begin{array}{c}
a_{\mathbf{k},\uparrow} \\
a_{-\mathbf{k},\downarrow} \\
a_{-\mathbf{k},\downarrow}^\dag \\
-a_{\mathbf{k},\uparrow}^\dag
\end{array}
\right),
\end{equation}
where for convenience, we have assumed a gauge in which the Bogoliubov
coefficients are real.
For a BCS superconductor with electronic dispersion $\varepsilon_{\bf k}$, 
gap function $\Delta_{\bf k}$ and quasiparticle energies 
$E_{\bf k} = \sqrt{\varepsilon_{\bf k}^2 + \Delta_{\bf k}^2}$ the Bogoliubov 
coefficients are
\begin{align}
u_{\mathbf{k}} & = \sqrt{\frac{1}{2} \left( 1 + 
\frac{\varepsilon_{\bf k}}{E_{\bf k}} \right)}, \\
v_{\mathbf{k}} & = \mbox{sign}(\Delta_{\bf k}) \sqrt{\frac{1}{2} 
\left( 1 - \frac{\varepsilon_{\bf k}}{E_{\bf k}} \right)}.
\label{v_k}
\end{align}

Introducing a Nambu spinor for the quasiparticle operators 
$\Phi_{\bf k}^\dag = ( a_{\mathbf{k},\uparrow}^\dag \, a_{-\mathbf{k},\downarrow}^\dag \, 
a_{-\mathbf{k},\downarrow} \, -a_{\mathbf{k},\uparrow} )$,  we may 
write the Bogoliubov transformation \eqref{Bogoliubov} as
\begin{equation}
\label{BogolNambu}
\Psi_{\bf k} = \left(u_{\mathbf{k}} \, \hat{1} + i \, v_{\mathbf{k}} \, \hat{\tau}_2 \right) 
\Phi_{\bf k},
\end{equation}
where $\hat{1}$ is the $4\times 4$ unit matrix and $\hat{\tau}_2$ is
the isospin matrix, 
\begin{equation}
\label{tau_2}
\hat{\tau}_2 = i\left(
\begin{array}{cr}
\underline{0} & -\underline{1} \\
\underline{1} & \underline{0} 
\end{array}
\right),
\end{equation}
where an  underscore denotes a two dimensional matrix. Using Eqs.~\eqref{scparticles} and 
\eqref{BogolNambu} the $t$-matrix can be written in the form 
$\hat  T = \sum_{\bk ,\bk '}\Phi_{\bk}^\dag \, \hat{T}^{QP}_{\bk ,\bk
'} \, \Phi_{\bf k'}$, where 
\begin{equation}
\label{T_QP}
\hat{T}^{QP}_{\bk ,\bk '} = 
\left(u_{\mathbf{k}} \, \hat{1} - i \, v_{\mathbf{k}} \, \hat{\tau}_2 \right)
\hat{T}
\left(u_{\mathbf{k}'} \, \hat{1} + i \, v_{\mathbf{k}'} \, \hat{\tau}_2 \right).
\end{equation}
Equation~\eqref{T_QP} 
shows that the scattering matrix elements of $\hat{T}^{QP}_{\bk ,\bk '}$ 
of the SC quasiparticles are determined by combinations of the 
Bogoliubov coefficients $u_{\mathbf{k}}$ and $v_{\mathbf{k}}$. 
Note that the sign of the SC order parameter $\Delta_{\bf k}$ enters the coherence 
factors according to Eq.~\eqref{v_k} so the momentum dependence of
quasiparticle scattering is sensitive to the way the 
phase of $\Delta_{\bf k}$ changes in momentum space. 

Now the action of the $\hat \tau_{2}$ matrix on the $t$-matrix is
closely related to its transformation under time reversal. The
time-reversed $t$-matrix $\hat T^{\theta }_{\bk ,\bk '}$ is determined
from its transpose in the following way: 
\begin{equation}\label{}
\hat T^{\theta }_{\bk ,\bk '} (\omega)
= 
\hat \sigma_{2}\hat T^{T}_{-\bk' ,-\bk } (-\omega) \hat \sigma_{2}, 
\end{equation}
where for generality, we have included the frequency dependence of
the $t$-matrix.
Now since the Nambu spinors satisfy the relation $\Psi_{-\bk }^{*}=(\Psi_{-\bk }\dg
)^{T}= \hat \sigma_{2}\hat \tau_{2}\Psi_{\bk }$, it follows from Eq.~(2) that an
arbitrary $t$-matrix satisfies the ``CPT'' relation
\[
\hat \tau_{2} \, \hat \sigma_{2} \, \hat T^{T}_{-\bk' ,-\bk } (-\omega) \, \hat \sigma_{2} \,
 \hat \tau_{2} = - \hat T_{\bk ,\bk '} (\omega).
\]

Combining these two relations, it follows that the time-reversed
$t$-matrix is given by
\[
\hat T^{\theta }_{\bk ,\bk '} (\omega)= 
- \hat \tau_{2} \, \hat T_{\bk ,\bk '} (\omega) \, \tau_{2},
\]
so that if the $t$-matrix has a well-defined parity $\theta $ under
time-reversal $\hat T^{\theta }_{\bk ,\bk '}  = 
\theta \hat T_{\bk ,\bk'}
$,  it follows that 
\[
\theta \hat T_{\bk ,\bk'} (\omega)
= 
- \hat \tau_{2} \, \hat T_{\bk ,\bk '} (\omega) \, \hat \tau_{2}.
\]
With this relationship, we can rewrite the quasiparticle $t$-matrix in
the form
\begin{equation}
\begin{split}
\label{T_QP1}
\hat{T}^{QP}_{\bk ,\bk '} & =
(u_{\bk }u_{\bk '}- \theta v_{\bk }v_{\bk '}) \hat T_{\bk ,\bk '} \\
& + (v_{\bk}u_{\bk '}+\theta v_{\bk '}u_{\bk}) (-i\hat \tau_{2}) \hat T_{\bk ,\bk '}.
\end{split}
\end{equation}
It is the {\sl diagonal elements} of this matrix that determine
quasiparticle scattering. 
If we restrict our attention to scattering processes that preserve
particle number, then the second term in this expression is
off-diagonal in the quasiparticle basis, and does not contribute to
low-energy quasiparticle scattering. In this case, the quasiparticle scattering
matrix is determined by the diagonal elements of 
\begin{eqnarray}\label{l}
\hat{T}^{QP}_{\bk ,\bk '} &=&
(u_{\bk }u_{\bk '}- \theta v_{\bk }v_{\bk '}) \hat T_{\bk ,\bk '}.
\\
&&(\hbox{particle-conserving scattering})\nonumber
\end{eqnarray}
Near the Fermi surface, the magnitudes of the Bogoliubov coefficients
are equal $|u_{\bk_{F}}|=|v_{\bk_{F}}|=\frac{1}{\sqrt{2}}$, so that
near the Fermi surface,
\begin{equation}\label{}
\hat{T}^{QP}_{\bk ,\bk '} \propto (1 - \theta {\rm sgn} (\Delta_{\bk }\Delta_{\bk '}))
\end{equation}
In other words, time-reverse even scattering ($\theta =1$),  gives
rise to sign-reversing scattering, while time-reverse odd scattering
($\theta =-1$) gives rise to sign-preserving scattering. 

Cases of particular interest are potential, magnetic and resonant scattering,
respectively denoted by $t$-matrices of the form
$\hat{\tau}_3$,
 $\hat{\sigma}_3$, and $\hat{1}\times t_{odd} (\omega)$, (
where $t_{odd} (\omega)=-t_{odd} (-\omega)$ denotes an odd function of
frequency).
The 
coherence factors and time-reversal parities corresponding to these
scattering mechanisms are listed 
in Table~\ref{table}. 
The application of a  magnetic field selectively enhances
time-reversal odd scattering.
We now describe a theoretical model for 
the field-dependent scattering of quasiparticles in an 
$
s_{\pm}$
superconductor. 

\begin{widetext}
\begin{center}
\begin{table}
\begin{tabular}[t]{ccccc}
\hline \hline $t$-matrix \quad & \quad Scatterer \quad & \quad 
Coherence factor \quad & \quad Enhances  \quad
&  \quad Time reversal \\ \hline
$\hat{\tau}_3$ & \quad Potential \quad & 
\quad $u_{\bf k} u_{\bf k'} - v_{\bf k} v_{\bf k'}$  \quad
&  \quad $+-$  \quad & \quad even \quad \\ 
$\hat{\sigma}_3$ & \quad Magnetic \quad & 
\quad $u_{\bf k} u_{\bf k'} + v_{\bf k} v_{\bf k'}$  \quad
&  \quad $++$  \quad & \quad odd \quad \\
$\hat{1}\times t_{odd} (\omega)$ & \quad Resonant \quad &  
\quad $u_{\bf k} u_{\bf k'} + v_{\bf k} v_{\bf k'}$  \quad
&  \quad $++$  \quad & \quad odd  \quad \\ \hline \hline
\end{tabular}
\caption{Relation between coherence factors and time reversal symmetry properties 
for a set of particle conserving scatterers. In resonant scattering,
the $t$-matrix is an  odd-function of frequency $t_{odd} (\omega)= -t_{odd}(-\omega).$
}
\label{table}
\end{table}
\end{center}
\end{widetext}

\subsection{BCS model of a Pauli-limited $s_{\pm }$ superconductor}

\begin{figure}[ht] \begin{center} \scalebox{0.3}{ \vspace{-2cm}
\includegraphics*{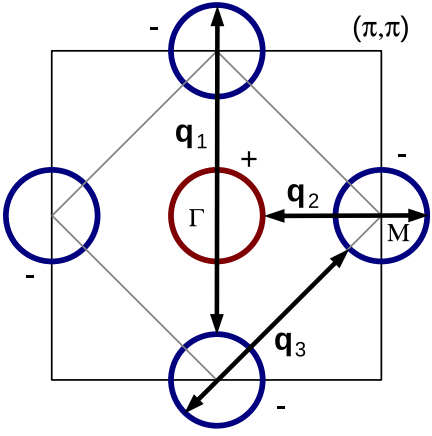} } \end{center} \caption{Schematic picture
of the disconnected Fermi surface of an iron-based superconductor. For
simplicity, two hole cylinders and two electron cylinders around
$\Gamma$ and M points are represented as circles. Signs of the SC gap
expected for $s_\pm$-wave symmetry are denoted as '+' and '-' in the
picture. Arrows denote inter-Fermi-pocket scatterings. The wavevectors ${\bf
q}_1$  and ${\bf q} _3$
are ``sign-preserving'', connecting pockets with the same
superconducting phase, while ${\bf q}_2$  is ``sign-reversing'', 
connecting pockets with opposite
superconducting phase. 
}  \label{Fig_FS} \end{figure}

The Fermi surface of iron-based superconductors consists of
two-dimensional hole and electron pockets centered at $\Gamma$ and M
points, respectively.\cite{SD2008} A schematic picture of the
disconnected Fermi surface is shown in Fig.~\ref{Fig_FS}. Since the
hole and the electron pockets are similar in both shape and volume,
the interband nesting between these pockets may generate spin
fluctuations at momentum ${\bf Q} = (\pi,0)$ which can be detected as
a coherence peak at ${\bf Q}$ in the Fourier-transformed local density
of states. If such a spin fluctuation induces the electron pairing, a
sign reversal of the SC-gap function occurs between the hole and
electron pockets, resulting in $s_{\pm}$-wave
symmetry.\cite{MS2008}-\cite{CT2009}

We employ a simple
two-band tight binding model with a singlet pairing.
In our model, we assume an extreme Pauli-limited superconductor, in
which the dominant effects of the applied magnetic field enter through
the Zeeman splitting of the conduction sea.  In our simplified
treatment, we assume that these effects are strong enough to permit us
to neglect orbital effects of the field. 

The model Hamiltonian is then
\begin{equation}
\label{BCSmodel}
\begin{split}
{\cal H} & = \sum_{{\bf k},\sigma} \left( \varepsilon^c_{{\bf k},\sigma} \, c_{{\bf k},\sigma}^\dag 
c_{{\bf k},\sigma} + \varepsilon^d_{{\bf k},\sigma} \, 
d_{{\bf k},\sigma}^\dag d_{{\bf k},\sigma} \right) \\
& - \sum_{\bf k} \left[ \Delta_{\bf k} \left( c_{{\bf k},\uparrow}^\dag c_{-{\bf k},\downarrow}^\dag 
+ d_{{\bf k},\uparrow}^\dag d_{-{\bf k},\downarrow}^\dag \right) + \mbox{h.c.} \right],
\end{split}
\end{equation}
where '$c$' and '$d$' denote the two electron bands describing an electron pocket at M point 
and a hole pocket at $\Gamma$ point, respectively. We model the Fermi
surface with the following  
dispersion  
\begin{align}
\varepsilon^c_{\mathbf{k},\sigma} & = 2t \left[ \mbox{cos}(k_x + k_y) 
+ \mbox{cos}(k_x - k_y) \right] + \mu_c - 
\sigma \mu_{B}B \nonumber \\ 
& \mbox{(electron pocket at M point)},  \\
\varepsilon^d_{\mathbf{k},\sigma} & = 2t 
\left( \mbox{cos}k_x + \mbox{cos}k_y \right) - \mu_d - \sigma \mu_{B}B\\
& \mbox{(hole pocket at $\Gamma$ point)}, \nonumber
\end{align}
where $\mu_{B}$ is the Bohr Magneton.
Note that here, all orbital
effects of the magnetic field on the dispersion and gap function
have been neglected. 
Finally, we assume a superconducting field of $s_{\pm}$-wave 
symmetry described by the gap function
\begin{eqnarray}
\Delta_{\mathbf{k}} &=& \tilde{\Delta} \left[ \mbox{cos}(k_x + k_y) +
\mbox{cos}(k_x - k_y)\right]\cr  &=& 2 
\tilde{\Delta} \cos k_{x}\cos  k_{y}.
\end{eqnarray}
In the numerical work reported here, we  took values
$\mu_c = 3.56t$, $\mu_d = 3.2t$, and $\tilde{\Delta} = 0.2t$.
With these parameters, the minimum value quasiparticle 
gap on the Fermi surface is given by
$\Delta  = 1.6\tilde{\Delta }= 0.32t$.

\section{Tunneling conductance}

Following the ideas of Ref.\cite{MC2009} in this section we 
numerically model the conductance ratio measured by an STM 
experiment\cite{FK2007}.
First, we  review the relation
between tunneling conductance and the local density
of states (LDOS). Within a simplified model of the tunneling process, the differential tunneling 
conductance $dI / dV (\mathbf{r},V)$ at a location $\mathbf{r}$ and
Voltage $V$ is given by
\begin{equation}
\label{Z_rho}
\frac{dI}{dV} (\mathbf{r},V) \propto |M(\mathbf{r})|^2 \rho(\mathbf{r},eV),
\end{equation}
where $\rho(\mathbf{r},\omega) = A(\mathbf{r},\mathbf{r},\omega)$ is the single-particle density 
of states at energy $\omega$ and $M(\mathbf{r})$ is the spatially 
dependent tunneling-matrix element, which includes contributions of the sample wave function around the 
tip. Thus, the tunneling conductance measures the thermally smeared LDOS of the sample 
at the position $\mathbf{r}$ of the tip. Note that a spatial
dependence of the LDOS often arises from a scattering off impurities distributed in the sample.

To filter out the spatial variations in the tunneling-matrix elements $M(\mathbf{r})$, the conductance 
ratio is taken as
\begin{equation}
\label{Z(r,V)}
Z(\mathbf{r},V) = \frac{\frac{dI}{dV} (\mathbf{r},+V)}{\frac{dI}{dV} (\mathbf{r},-V)} = \frac{\rho_0(eV) + 
\delta \rho(\mathbf{r},eV)}{\rho_0(-eV) + \delta \rho(\mathbf{r},-eV)}.
\end{equation}
The first part $\rho_0(\pm eV)$ of the LDOS describes the sample-averaged tunneling density of states 
at bias voltage $\pm V$. The spatial fluctuations of $\rho(\mathbf{r},\pm eV)$ are given by the second part 
$\delta \rho(\mathbf{r},\pm eV)$. 
For small fluctuations $\delta \rho(\mathbf{r},\pm eV) \ll \rho_0 (\pm eV)$
the Fourier transform of expression \eqref{Z(r,V)} is 
\begin{equation}
\label{Z(q,V)}
\begin{split}
Z(\mathbf{q},V) & = Z_0(V) (2\pi)^2 \delta(\mathbf{q}) \\
& + Z_0(V) 
\left[ \frac{\delta \rho(\mathbf{q},eV)}{\rho_0(eV)} - \frac{\delta \rho(\mathbf{q},-eV)}{\rho_0(-eV)}
\right],\\
&
\end{split}
\end{equation}
where $Z_0(V) = \frac{\rho_0(eV)}{\rho_0(-eV)}$. The Fourier transformed conductance ratio 
$Z(\mathbf{q},V)$ consists of a single delta function term 
at $\mathbf{q} =0$ and a backround, which describes the interference patterns
produced by quasiparticle scattering off impurities. The condition for small fluctuations
is satisfied in the clean limit 
at finite and sufficiently large bias voltages $|V| > 0$.\cite{MC2009}

It proves useful to write the fluctuations in the conductance ratio as a sum of two terms,
even and odd in the bias voltage
\begin{equation}
\begin{split}
\label{Z_odd_even}
Z(\mathbf{q},V)|_{\mathbf{q} \ne 0} &= Z_0(V) \left\{ \delta \rho^{+}(\mathbf{q},eV) \left[ 
\frac{1}{\rho_0(eV)} - \frac{1}{\rho_0(-eV)} \right] \right.\\
&+ \left. \delta \rho^{-}(\mathbf{q},eV) \left[ 
\frac{1}{\rho_0(eV)} + \frac{1}{\rho_0(-eV)} \right] \right\},
\end{split}
\end{equation}
where $\delta \rho^{\pm}(\mathbf{q},\omega) = 
[\delta \rho(\mathbf{q},+\omega) \pm \delta \rho(\mathbf{q},-\omega)]/2$. Depending on the particle-hole 
symmetry properties of the sample-averaged tunneling density of states $\rho_0(V)$, one of these 
terms can dominate.

We calculate the fluctuations of the LDOS in Eq.~\eqref{Z_odd_even} using the Green's function 
of a superconductor 
in the presence of impurities. As is discussed in Sec.~II, the electron field inside a superconductor
can be described by a four-component vector which is written as
\begin{equation}
\label{spinor_r}
\Psi^\dag(\mathbf{r},\tau) =
\left(
\psi_{\uparrow}^\dag ({\bf r},\tau ), 
\psi_{\downarrow}^\dag({\bf r},\tau ),
\psi_{\downarrow}({\bf r},\tau ), 
-\psi_{\uparrow}({\bf r},\tau )
\right) 
\end{equation}
in real space, where $\mathbf{r}$ denotes the position vector 
and $\tau$ is imaginary time. 
The matrix Green's function is defined as the time-ordered average,
\begin{equation}
\label{2}
\hat{G}_{\alpha\beta}(\mathbf{r}',\mathbf{r};\tau) = -\langle T_\tau \Psi_{\alpha}(\mathbf{r}',\tau)
\Psi_{\beta}^\dag(\mathbf{r},0)\rangle.
\end{equation}
For an electronic system in the presence of impurities, the Green's function \eqref{2} is usually 
calculated using the $t$-matrix method.\cite{BVZ2006,HVW1986} As is already discussed above, for a single impurity 
with scattering potential $\hat{U}$, 
the $t$-matrix exactly accounts for multiple scattering off that impurity. In momentum space, the 
$t$-matrix  is determined by the following self-consistent 
equation:
\begin{equation}
\label{t_series}
\begin{split}
\hat{T}(\mathbf{k},\mathbf{k}',\omega) & =  \hat{U}(\mathbf{k},\mathbf{k}') \\
& + \sum_{\mathbf{k}''}
\hat{U}(\mathbf{k},\mathbf{k}'') \hat{G}_0(\mathbf{k}'',\omega) \hat{T}(\mathbf{k}'',\mathbf{k}',\omega),
\end{split}
\end{equation}
where $\hat{U}(\mathbf{k},\mathbf{k}')$ is the scattering potential of a single impurity and 
$\hat{G}_0(\mathbf{k},\omega)$ is the bare Green's function of the BCS superconductor without 
impurities. The Green's function for an electron with a normal-state dispersion $\varepsilon_{\bf k}$ 
and gap function $\Delta_{\bf k}$ in an external magnetic field $B$ is
\begin{equation}
\label{G_BCS}
\hat{G}_0(\mathbf{k},\omega) = [ \omega 
\hat{1} - \varepsilon_{\mathbf{k}} \hat{\tau}_3 - 
B \hat{\sigma}_3 - \Delta_{\mathbf{k}} \hat{\tau}_1 ]^{-1}.
\end{equation}

Using $\hat{G}_0(\mathbf{k},\omega)$ and Eq.~\eqref{t_series},
the Fourier transformed Green's function can be 
written as
\begin{equation}
\label{G_fluct}
\hat{G}(\mathbf{k},\mathbf{k}',\omega) = \hat{G}_0(\mathbf{k},\omega) + \hat{G}_0(\mathbf{k},\omega)
\hat{T}(\mathbf{k},\mathbf{k}',\omega) \hat{G}_0(\mathbf{k}',\omega).
\end{equation}
Note that since translational invariance is broken by impurities, the Green's function depends on 
two momenta, ${\bf k}$ and ${\bf k}'$.

The LDOS is determined by the analytic continuation $\hat{G}(\mathbf{r}',\mathbf{r};i\omega_n) 
\rightarrow \hat{G}(\mathbf{r}',\mathbf{r};z)$ of the Matsubara Green's function
$\hat{G}(\mathbf{r}',\mathbf{r};i\omega_n) = \int_0^\beta \hat{G}(\mathbf{r}',\mathbf{r};\tau) 
e^{i(2n + 1) \pi T \tau} d\tau $,
\begin{equation}
\label{LDOS_r}
\rho(\mathbf{r},\omega) = \frac{1}{\pi} \mbox{Im} \mbox{Tr} \frac{\hat{1} + \hat{\tau}_3}{2} 
\left[\hat{G}(\mathbf{r},\mathbf{r}; \omega - i \delta)\right].
\end{equation}

Using Eqs.~\eqref{LDOS_r} and \eqref{G_fluct}, 
the Fourier transformed  
fluctuations $\delta \rho^{\pm} (\mathbf{q},\omega)$ in 
Eq.~\eqref{Z_odd_even} can be written in the compact form\cite{MC2009}
\begin{equation}
\label{rho_odd}
\begin{split}
\delta \rho^{\pm} (\mathbf{q},\omega) & = \\
\frac{1}{2\pi} \mbox{Im} \sum_{\mathbf{k}} & \mbox{Tr}
[\left(\begin{array}{c}1\cr \tau_{3}\end{array} \right)
 \, \hat{G}_0 (\mathbf{k}_- ,z)\, 
\hat{T} (\mathbf{k}_-,\mathbf{k}_+,z) \,
\hat{G}_0 (\mathbf{k}_+ ,z)],
\end{split}
\end{equation}
where $\mathbf{k}_{\pm} = \mathbf{k} \pm \frac{\mathbf{q}}{2}$ and $z = \omega - i \delta$. 
The ${\bf q}$ dependent fluctuation of the LDOS is a probe of all scattering processes 
combining two momenta ${\bf k}_+$ and ${\bf k}_-$ with a fixed
difference ${\bf q}$.

The only place that a magnetic field enters into the theory, is
inside the bare Green-functions, Eq.~(\ref{G_BCS}); the 
effect of the Zeeman splitting on the Green's functions
and the $t$-matrices can be simply understood as a result of 
making the substitution
\begin{equation}\label{}
\omega \longrightarrow \omega  - \hat \sigma_{3} \mu_{B}B.
\end{equation}
This substitution will, in general, modify both
the Green's functions  and the $t$-matrix.
Therefore, the breaking of time reversal symmetry can affect the system in two
ways: (i) a polarization of the Fermi surface, i.e., a different chemical potential 
for spin up and down electrons 
and  (ii) for an energy dependent
$t$-matrix (resonant scatterer), a magnetic field will 
lead to a difference in the scattering amplitudes 
of spin up and down electrons. 

\section{Results}

We numerically evaluated 
$\delta \rho^{\pm} (\mathbf{q},\omega)$ from Eq.~\eqref{rho_odd} using
the model \eqref{BCSmodel}, considering various different types of impurities. 
For simplicity, we assumed that 
the scattering is equal for the two types of electrons '$c$' and
'$d$'.

We computed the Fourier-transformed conductance 
ratio $Z(\mathbf{q},V)|_{\mathbf{q} \ne 0}$ 
from Eq.~\eqref{Z(q,V)} by evaluating numerically  the fluctuations 
$\delta \rho^{\pm} (\mathbf{q},\omega)$ from Eq.~\eqref{rho_odd} for potential, 
magnetic, and resonant scattering. 
The different types of scattering were modelled 
using $t$-matrices of the form listed in Table~\ref{table}.  The
computed influence of a magnetic 
field is illustrated in Figs.~\ref{Fig_1}-\ref{Fig_3}, where $Z (\mathbf{q},V)$
is plotted as a function of momentum $\mathbf{q}$ for 
two different values of the magnetic field $B$ at a fixed bias
voltage $eV=\Delta /2$, where $\Delta $ is the magnitude of the gap on
the electron and hole Fermi surface pockets. ($\Delta = 0.32
t$ in our numerical calculations.) The two
chosen values of $B$ are $\mu_B B = \Delta$ (upper panels) and
$\mu_B B = \Delta/2$ (lower panels).
The voltage $V$ has been chosen to coincide with the bottom of the
quasiparticle gap  at the lower field value. 

In each of Figs.~\ref{Fig_1}-\ref{Fig_3},
the value of $Z$ is normalized with respect to its maximum.
Furthermore, for a numerical evaluation of the sum over ${\bf k}$ in Eq.~\eqref{rho_odd}, 
we discretized 
the Brillouin zone in $N = 10000$ $\mathbf{k}$ points.

\subsection{Potential scattering}

At first, let us consider the case where the conduction electrons 
and non-magnetic impurity atoms interact each other via a Coulomb potential.
Assuming the Coulomb interaction is screened at 
length scales comparable to the lattice spacing, we consider a local
 scattering potential. For a simple modeling of the scattering, we apply Born approximation, 
which is equivalent to taking 
only the first term in Eq.~\eqref{t_series}. The $t$-matrix used in the calculations is
\begin{equation}
\hat{T}_P = \hat{\tau}_3.
\end{equation}

\begin{figure}[hbt]
      \includegraphics[width=0.7\linewidth]{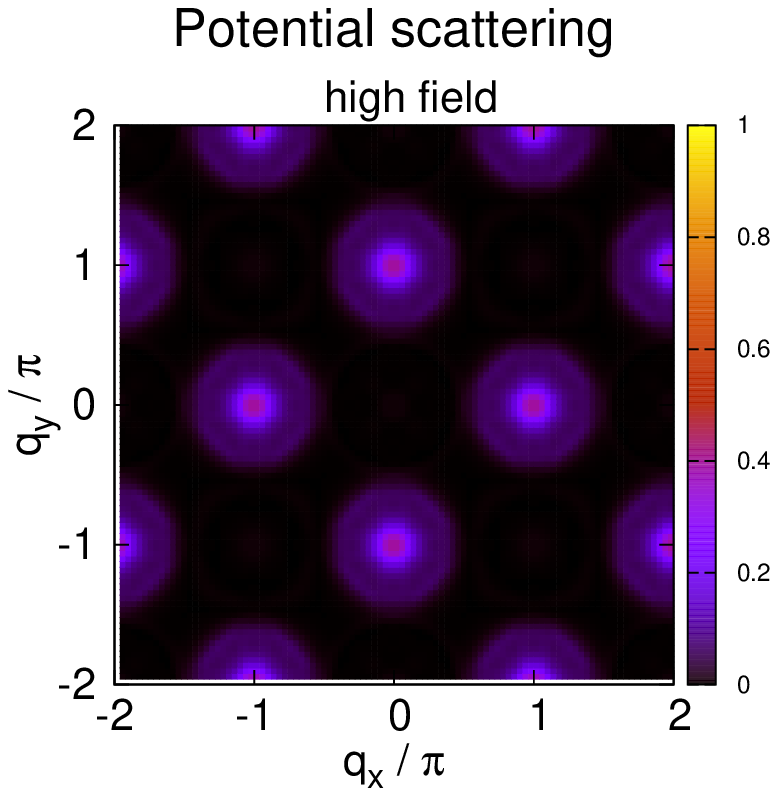} \vspace{-0.5cm}\\
      \includegraphics[width=0.7\linewidth]{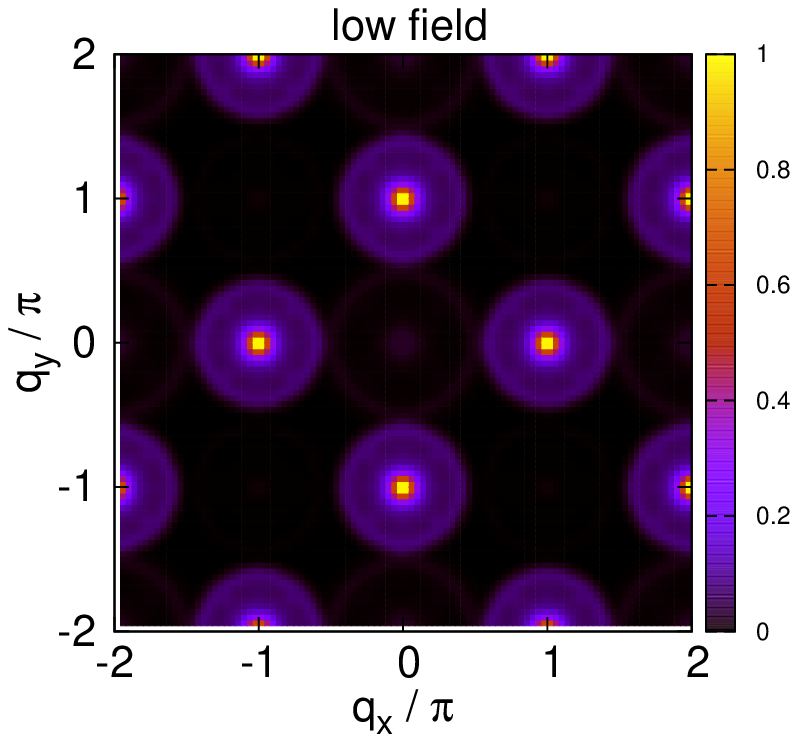}
  \caption{
Normalized conductance ratio $Z(\mathbf{q},V)|_{\mathbf{q}
  \ne 0}$ produced by potential scattering, 
evaluated at fixed bias voltage $eV = \Delta/2 $, at ``high 
field'' $\mu_{B}B = \Delta$ (upper panel) and ``low field'' $\mu_{B}B
= \Delta / 2$ (lower panel). Potential scattering induces
predominantly sign-reversing scattering ($\bf{q}= {\bf{ q}}_{2}$),
which is suppressed by a field.
}
   \label{Fig_1}
  \end{figure}

Figure~\ref{Fig_1} shows the conductance ratio  as a function 
of wave vector $\mathbf{q}$. We obtain a large quasiparticle 
interference at the $\mathbf{q}$-vectors $\mathbf{q} = (0,\pm \pi)$ and $\mathbf{q} = (\pm \pi,0)$ 
(denoted by ${\bf q}_2$ in Fig.~\ref{Fig_FS})
connecting Fermi pockets with opposite sign of SC gap (sign-reversing scattering). The signals 
for the sign-preserving 
$\mathbf{q}$-vectors $\mathbf{q} = (\pm \pi,\pm \pi)$ and $(0,0)$ (denoted by ${\bf q}_1$ and 
${\bf q}_3$ in Fig.~\ref{Fig_FS}) are  suppressed. The obtained sign change of the SC gap is a 
characteristic feature of potential scattering which can also be derived from its coherence 
factor $\left( u_{\bf k} u_{\bf k'} - v_{\bf k} v_{\bf k'} \right)$ (compare Table~\ref{table}). 

As is seen from the upper panel of Fig.~\ref{Fig_1}, a polarization of the 
Fermi surface by applying a magnetic field does not change the result qualitatively. The intensity 
of the sign-reversing peaks at ${\bf q}_2$ becomes somewhat smaller with increasing $B$. 
Note that although the Fermi surface is polarized by the magnetic field, a Coulomb potential is 
not sensitive with respect to the spin of scattered electrons.

\subsection{Magnetic scattering}

In addition to electrostatic interactions, if the impurity atom has 
a magnetic moment, there is an exchange interaction between the local spin on the impurity site and the spin of
conduction electrons. For the scattering model, we apply again Born approximation and 
assume a local exchange function. The corresponding $t$-matrix that is used in the numerics is
\begin{equation}
\hat{T}_M = \hat{\sigma}_3.
\end{equation}
\begin{figure}[hbt]
      \includegraphics[width=0.7\linewidth]{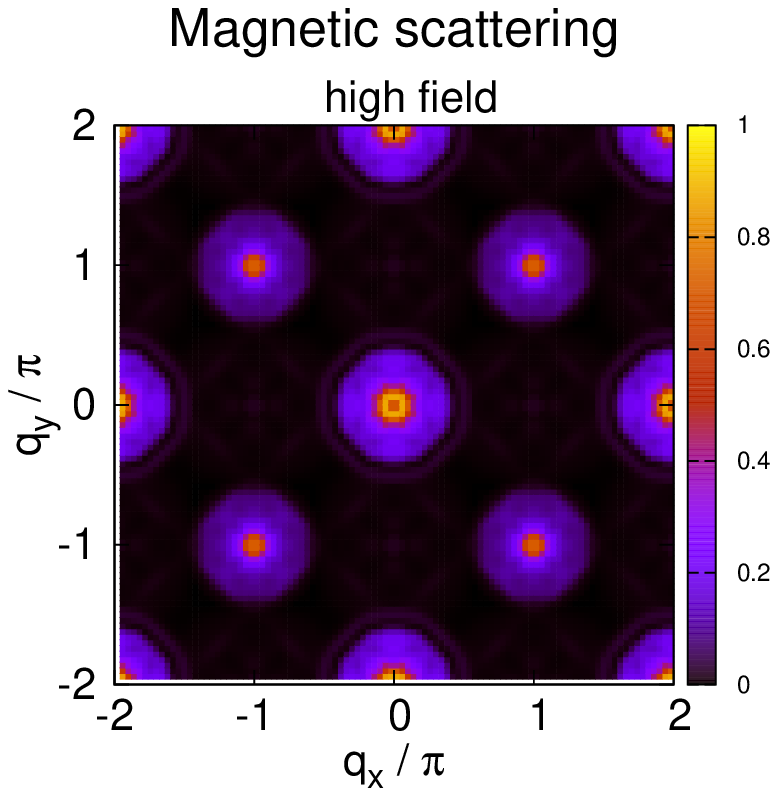} \vspace{-0.5cm}\\
      \includegraphics[width=0.7\linewidth]{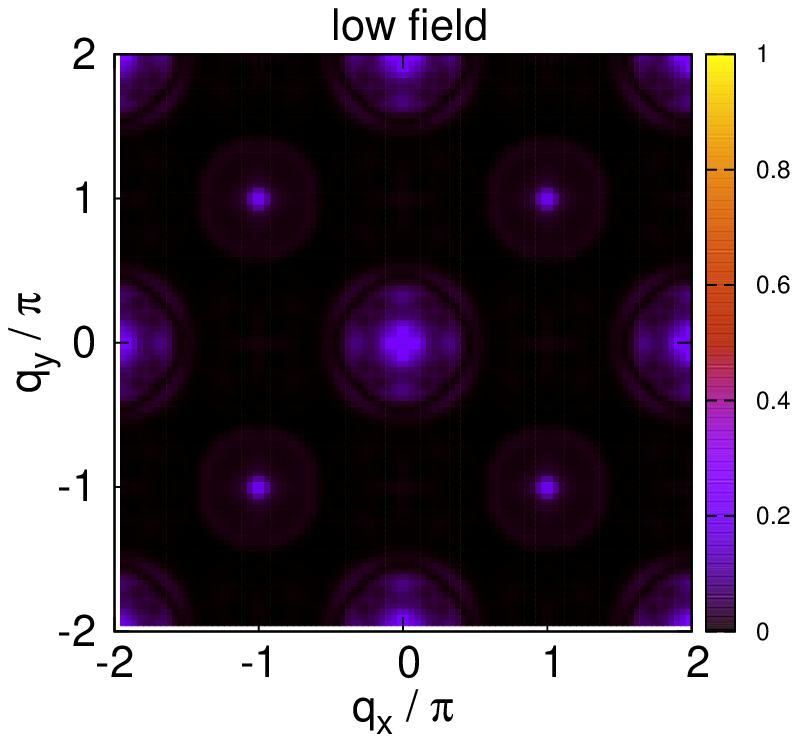}
  \caption{
Normalized conductance ratio $Z(\mathbf{q},V)|_{\mathbf{q}
  \ne 0}$ produced by magnetic scattering, 
evaluated at fixed bias voltage $eV = \Delta / 2 $, at ``high 
field'' $\mu_{B}B = \Delta$ (upper panel) and ``low field'' $\mu_{B}B
= \Delta / 2$ (lower panel). Magnetic scattering induces
predominantly sign-preserving scattering ($\bf{q}= {\bf{ q}}_{1,3}$),
which is {\sl enhanced } by a field.
Note that in Born approximation, there is no contribution 
from magnetic scatterers in the zero-field case.}
   \label{Fig_2}
  \end{figure} 
Figure~\ref{Fig_2} presents the conductance ratio for pure magnetic scatterers. 
The quasiparticle interference is enhanced for  
the sign-preserving scattering processes at ${\bf q}$ vectors ${\bf q}_1$ and ${\bf q}_3$ 
and the intensity of the peaks 
is almost equal. On the other hand, sign-reversing processes are suppressed.
Furthermore, we obtain  a characteristic  increase of  the peak intensity 
with increasing values of $B$. These results are consistent with other
theoretical works that are based on more realistic band models\cite{ZF2009},\cite{AK2010}. 
In the zero-field case, 
$B=0$ LDOS fluctuations coming from magnetic scatterers are suppressed completely. 
The reason is the proportionality of the
scattering potential  to the electron spin. Therefore, contributions coming
from spin up and down cancel each other when the Hamiltonian is symmetric under time reversal.
For $B>0$, where the time reversal symmetry is broken and spin up and down electrons have different 
chemical potential, coherence factors 
$\left( u_{\bf k} u_{\bf k'} + v_{\bf k} v_{\bf k'} \right)$ associated with a $t$-matrix 
proportional to $\sigma_3$ lead 
to sign-preserving scattering. Thus, for magnetic scatterers, the effect of a $B$ field is driven
by a Zeeman-splitting of the bulk superconductor.

\subsection{Resonant scattering}

Resonant scattering is a multiple scattering process, 
where the main contributions to the 
$t$-matrix derive from higher order terms in 
which the energy dependence is determined
by the bare Green's function $\hat{G}_0(\mathbf{k},\omega)$.
The effect of a magnetic field  is to 
replace
\[
\omega\longrightarrow \omega - \sigma_{3}\mu_{B}B,
\]
inside all Green's functions, and 
thus a resonant 
$t$-matrix 
in an external magnetic field $B$ can be related to the corresponding 
zero-field form, 
as follows
\begin{equation}
\hat{T}_R[\omega;B] = T_{R}[\omega \hat{1} - \mu_{B}B \hat{\sigma}_3 ].
\end{equation}
An expansion to linear order in the field 
\begin{equation}\label{expT}
\hat{T}_R(\omega;B) = T_{R}(\omega) \, \hat{1} - 
T_{R}'(\omega) \,B\, \hat{\sigma}_3 + \dots
\end{equation}
immediately shows that the leading correction to the zero-field scattering plays the same
role as a magnetic impurity. For a phenomenological treatment 
of resonant scattering, we have taken the form
\begin{eqnarray}\label{T_R}
\hat{T}_R(\omega;B) &=& \frac{1}{(\omega - i \Gamma) \hat{1} -\mu_{B} B
\hat{\sigma_3}}\cr &=& \frac{(\omega - i \Gamma) \hat{1} + \mu_{B}B\hat \sigma_{3}}{
(\omega - i\Gamma)^{2}- (\mu_B B)^{2}}
,
\end{eqnarray}
where we chose the width $\Gamma$ to be equal to the superconducting
gap $\Delta $ at the
Fermi-surface. 
\begin{figure}[hbt]
      \includegraphics[width=0.7\linewidth]{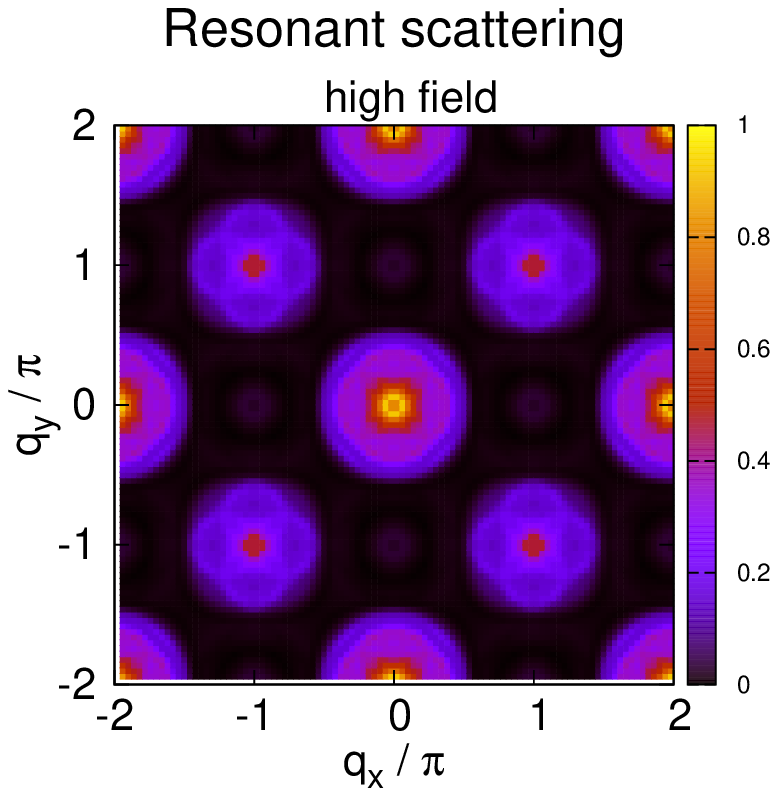} \vspace{-0.5cm}\\ 
      \includegraphics[width=0.7\linewidth]{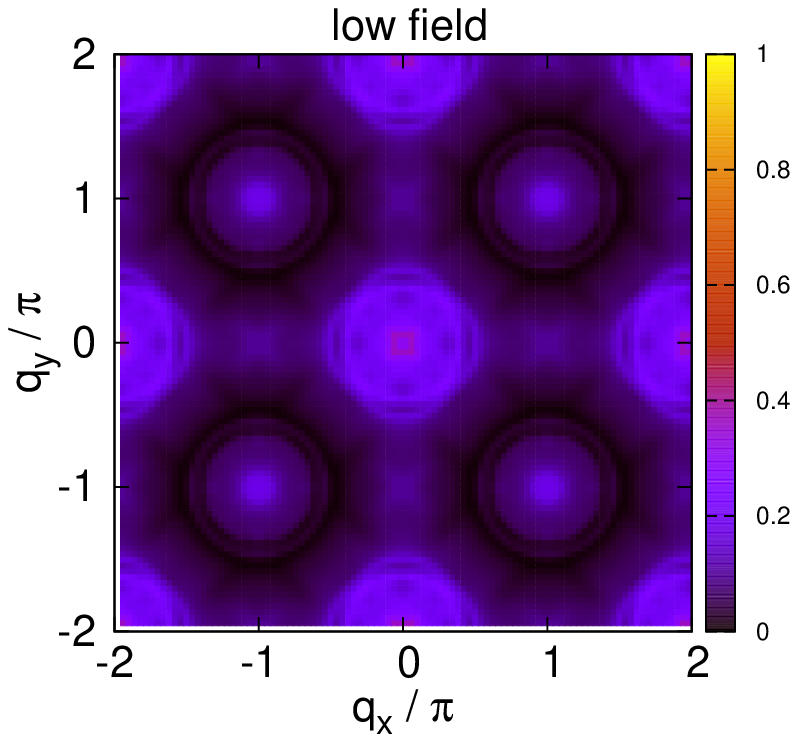}
  \caption{
Normalized conductance ratio $Z(\mathbf{q},V)|_{\mathbf{q}
  \ne 0}$ for resonant scattering, 
evaluated at fixed bias voltage $eV = \Delta/2 $, at ``high 
field'' $\mu_{B}B = \Delta$ (upper panel) and ``low field'' $\mu_{B}B
= \Delta / 2$ (lower panel). Increasing the field leads to an
enhancement of sign-preserving scattering 
($\bf{q}= {\bf{ q}}_{1,3}$)}
   \label{Fig_3}
  \end{figure}

The result for $Z(\mathbf{q},V)|_{\mathbf{q}
  \ne 0}$ is shown in Fig.~\ref{Fig_3}. In low field case
$\mu_{B}B = \Delta / 2$ (lower panel), we obtain  sign-preserving peaks
at ${\bf q} = (0,0)$ and $\mathbf{q} = (\pm \pi,\pm \pi)$ of rather weak intensity.
Moreover,  some slight structures around these peaks 
appear in the QPI.
Increasing the field leads to an enhancement of sign-preserving scattering which
can be seen clearly in the upper panel of Fig.~\ref{Fig_3} where the intensity of 
the sign-preserving peaks
at $\bf{q}= {\bf{ q}}_{1,3}$ is increased noticeably as compared to the lower 
field case. Moreover, sign-reversing contributions to the QPI become strongly suppressed 
for higher fields so that we obtain a  qualitatively similar picture as for pure magnetic scatterers 
(compare Fig.~\ref{Fig_2}). This behaviour is also expected from the $t$-matrix model~\eqref{T_R}
which predicts for large enough field values a dominant term proportional to $\hat{\sigma_3}$
leading to sign-preserving scattering.

\subsection{Discussion}

The results in Figs.~\ref{Fig_1}-\ref{Fig_3} show a characteristical finite width of 
$\Delta q \approx 0.2\pi$ for all QPI peaks. 
This property is determined by the nesting for scattering processes between 
the different Fermi surfaces 
(compare Fig.~\ref{Fig_FS}). With Ref.~\cite{Mazin} 
in mind we note that the sign preserving QPI peaks at ${\bf q}_3$ 
might be very sharp, since  the nesting condition is perfectly
fulfilled for intrapocket scattering. 

In the present work, we have neglected all orbital effects of the magnetic field
on the quasiparticle interference. In order to justify this approximation, we 
recalculated the conductance ratio for all kinds of scattering under the assumption that 
the magnetic field is now coupled to the degenerate $d_{xz}$ and $d_{yz}$ orbitals within 
a simplified two orbital model for iron-based superconductors.\cite{R2008} As a main 
result, we found that a magnetic field of the order of $\Delta$ which is coupled to the
orbitals leads to a slight deformation of the Fermi surface giving rise to
a small change of the nesting conditions. In all cases of scattering, the
orbital coupling affected rather the shape than the intensities of the quasiparticle 
interference peaks which justifies the assumption that the orbital effect in Pauli-limited
superconductors is very small in comparism to the field effect based on Zeeman splitting.

\section{Conclusions}

To summarize, we have modelled the effect of Zeeman splitting on
quasiparticle interference in iron-based superconductors using an 
$s_{\pm}$-wave-symmetry of the superconducting order
parameter. Our model is applicable under the assumption that
these superconductors are Pauli limited. 
We investigated three cases of
scattering: 
The effect of a spin-split spectrum on scattering off (i)
nonmagnetic potential scatterers, (ii) magnetic impurities, and (iii)
nonmagnetic resonant scatterers. While the field has almost no effect
on scattering off potential impurities, in the cases (ii) and (iii),
the main effect of the 
field is to enhance time reversal odd scattering, enhancing
the sign-preserving  points and depressing sign-reversing 
points in the QPI pattern. A
possible scattering model that would be in accordance with STM
experiments recently performed on the chalcogenide superconductor
Fe(Se,Te) consists of potential impurities and resonant
scatterers. However, so far there is no direct experimental evidence for the
presence of resonant scatterers in these fully gapped superconductors. 
Possible other
effects that arise from breaking the time reversal symmetry such as
the influence of a supercurrent on nonmagnetic scatterers is an
interesting open question that deserves future investigation.

\section*{Acknowledgments}
The authors would like to thank T.~Hanaguri and H.~Takagi for
discussions related to this work. 
This work was supported by DFG grant SY 131/1-1 (SS) and
DOE grant DE-FG02-99ER45790 (PC).


\end{document}